\documentclass[11pt,twoside]{article}

\usepackage{asp2006}
\usepackage{epsf}
\usepackage{psfig}
\usepackage{lscape}
\usepackage{amssymb}

\newcommand{\Mvar}{$M_{\rm BH}^{\sigma^2_{\rm nxs}}$}
\newcommand{\Mlit}{$M_{\rm BH}^{\rm lit}$}
\newcommand{\Gs}{$\Gamma_{\rm 0.1-2.4 keV}$}
\newcommand{\Hb}{H{\small$\beta$}}
\newcommand{\sig}{$\sigma^2_{\rm nxs}$}

\begin{document}
\title{Black hole masses in NLS1 galaxies from the X-ray excess
variance method}

\author{Marek Niko\l ajuk\altaffilmark{1}, Pawe\l\
Gurynowicz\altaffilmark{1}, and Bo\.zena Czerny\altaffilmark{2}}

\altaffiltext{1}{Institute of Theoretical Physics, University of 
Bia\l ystok, Lipowa 41, PL-15352 Bia\l ystok, Poland. Send requests 
to M.Nikolajuk: mrk@alpha.uwb.edu.pl}
\altaffiltext{2}{N. Copernicus Astronomical Center, Polish Academy of
Sciences, Bartycka 18, PL-00716 Warsaw, Poland}

\begin{abstract}

We estimate black hole masses in Narrow Line Seyfert 1 (NLS1) galaxies
at the basis of their X-ray excess variance. We apply the standard
approach appropriate for Broad Line Seyfert 1 (BLS1) galaxies. In
general, we find that the obtained masses are by a factor $\sim$ 20 too
small to agree with values obtained from other methods (reverberation,
stellar dispersion). However, a small subset of our
NLS1 objects does not require that multiplication, or the correction
factor is less than 4. We find that this subset have a soft X-ray
photon index, \Gs, smaller than 2. We thus postulate that this
subclass of NLS1 actually belongs to BLS1.
\end{abstract}

\section{Introduction}
NLS1 sources are defined as a subclass of broad line objects at the basis
of the optical spectra, as sources with the Full Width at Half Maximum 
(FWHM) of \Hb\ line, produced
in Broad Line Region, smaller than 2100
kms$^{-1}$. In BLS1s FWHM(\Hb) is $\gtrsim$ 2100 kms$^{-1}$. It was later
found that NLS1 
generally have smaller black hole masses for a given luminosity, and
strong soft X-ray excesses (Mathur et al. 2001; Crummy et al,
2006). Soft X-ray photon indices, \Gs, of NLS1 sources 
have a mean value 3.1, while BLS1
sources are harder, with a mean of 2.1 (e.g. Boller et
al. 1996). It suggests that NLS1 and BLS1 differ with respect to the 
accretion pattern.

We have considered a sample of 21 NLS1 galaxies (e.g. Mrk 110, 335,
478, Ark 564, NGC 4051, 5506, Ton S180, I Zw 1, PG 1211+143) and a
sample of BLS1 galaxies (e.g. NGC 4151, 5548, 7469, 3C 120, F9)
studied by Niko\l\/ajuk et al. 2006. All sources were monitored both
in optical and in X-ray band.

The masses of supermassive black hole for all our galaxies were taken
from literature, \Mlit. They were obtained with the reverberation
method (e.g. Peterson et al. 2004) or with the stellar velocity
dispersion method (e.g. Woo \& Urry 2002). Those masses were used to
compare with values obtained from X-ray variability. We have
determined the masses using the variance method described by
Niko\l\/ajuk et al. 2006. It is based on the scaling of the normalized
variance of X-ray light curves, \sig, measured in range 2-10 keV with
the $M_{\rm BH}$ value.

\section{Results and a small subset of NLS1 galaxies}

\begin{figure}
  \plottwo{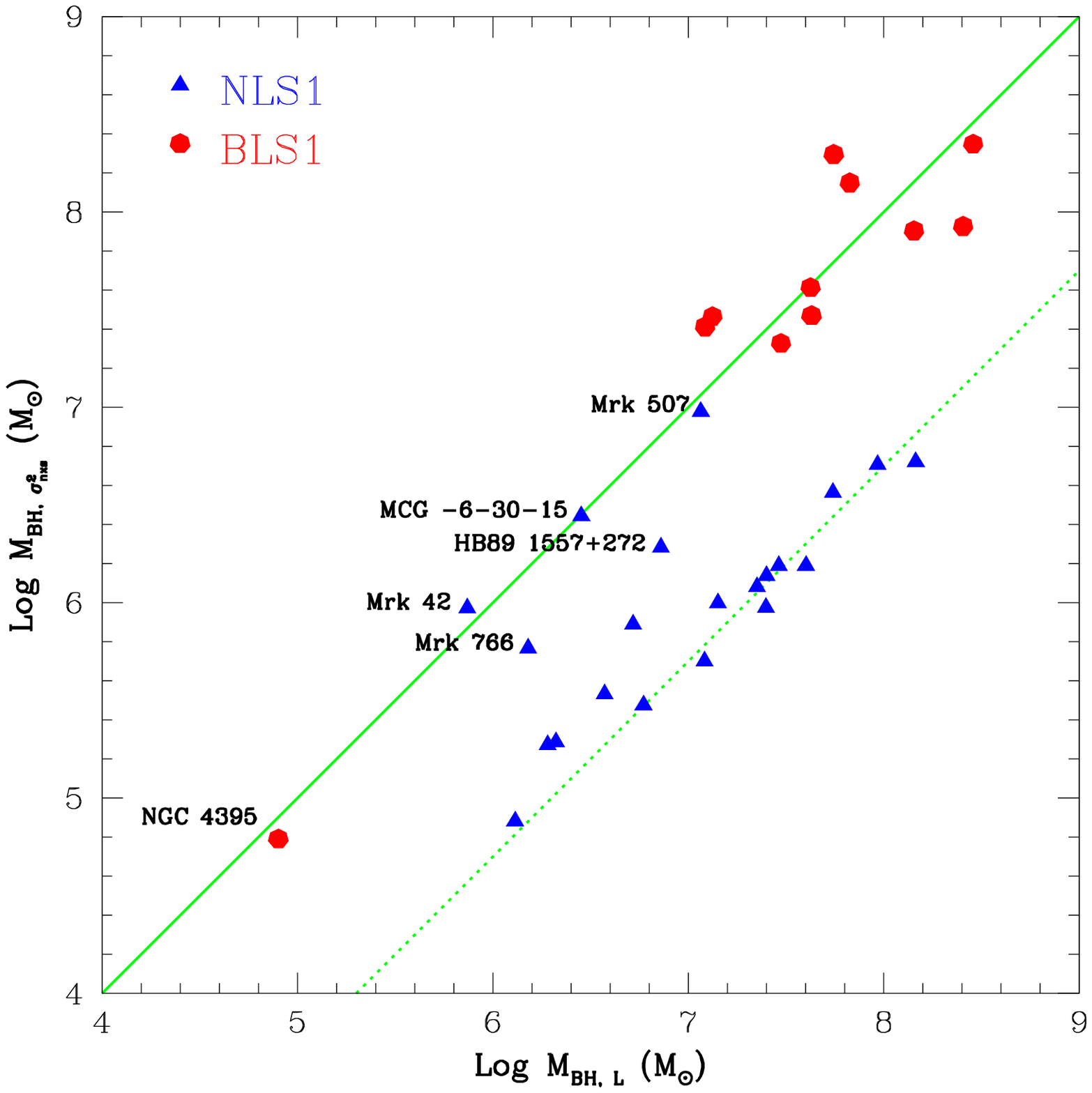}{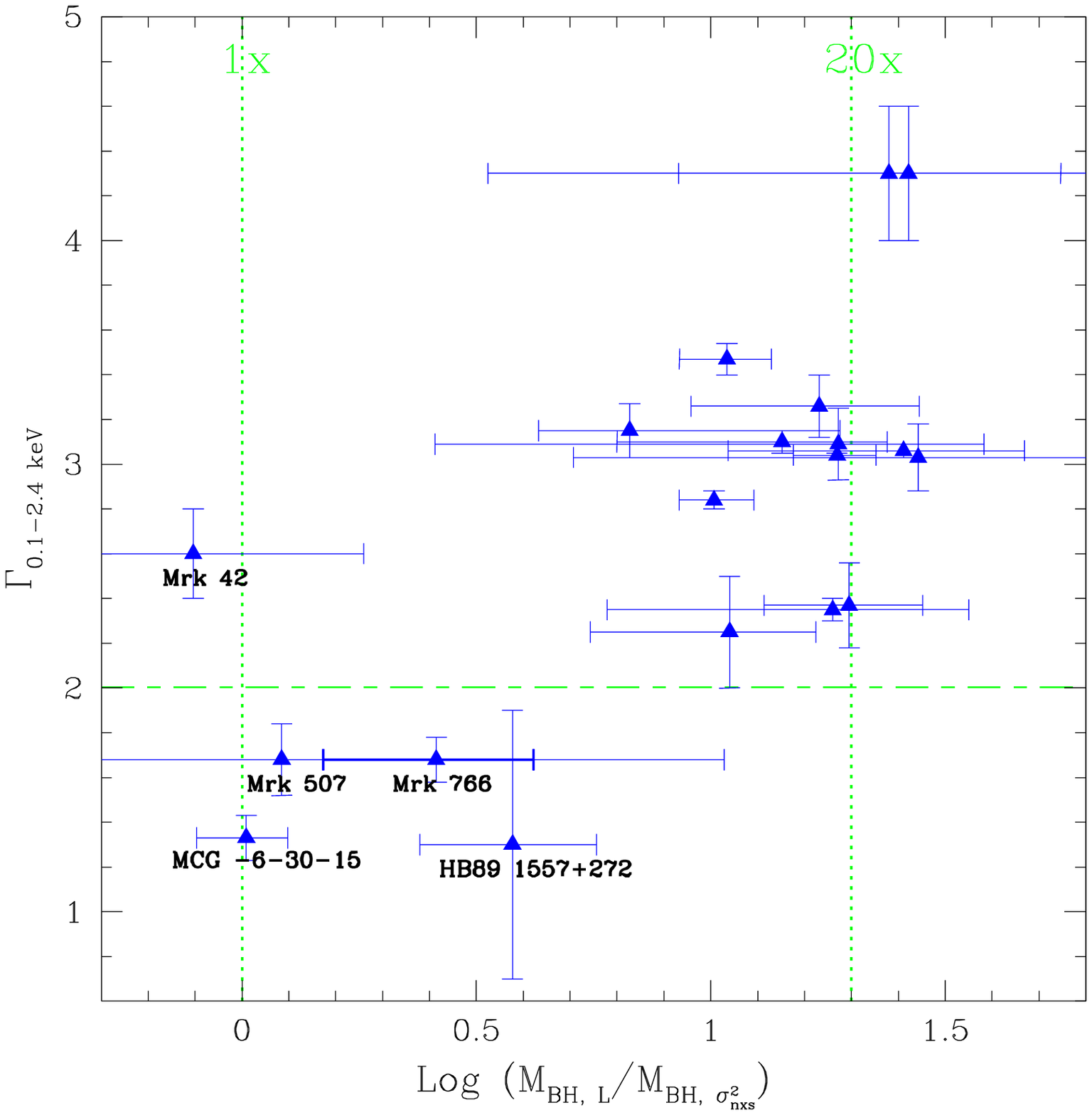} 
  \caption{Left panel:
  Black hole masses estimated with X-ray variance method versus masses
  taken from literature. The solid line shows the relation
  \Mlit=\Mvar. The dotted line shows the relation \Mlit=20\Mvar. Right
  panel: The soft X-ray photon index plotted versus the ratio
  \Mlit/\Mvar. The dotted lines show how many times \Mvar\ should be
  multiplicated in order to obtain \Mlit. The dashed line shows
  arbitrary border \Gs=2.  }
  \label{fig:MG}
\end{figure}

Generally, values of \Mvar\ require a multiplication by some
factor. Mean value of this factor is 20 (see e.g. Papadakis,
2004). Surprisingly, this factor for four galaxies (Mrk 42, Mrk 507,
Mrk 766, MCG -6-30-15) is smaller than 3 and for HB89 1557+272 is
smaller than 4 (Fig.~\ref{fig:MG}). The variability of
X-ray light in the 2-10 keV range of those 5 NLS1 galaxies is thus very
similar to all BLS1 sources.

We find that this subset have values of soft X-ray photon index,
\Gs, less than or equal to $\sim$ 2. Once again it is behaviour of
BLS1 galaxies. Spectra in the soft X-ray range of those 5 NLS1 are
similar to spectra of BLS1.

On the other hand when we compare accretion rates, we cannot see
difference between those 5 NLS1 and other NLS1 galaxies. This
small subset of NLS1 accrete, like other NLS1, with $\dot m=\dot
M/\dot M_{\rm Edd}$ higher than 0.1 or even close to Eddington
limit. This is opposite to BLS1 sources, which accrete with
$\dot m \lesssim 0.1$. 


%
%
%

\acknowledgements
This work was partially supported by the grant \break 1P03D00829 of the Polish State Committee for Scientific Research.


 \end{document}